\documentclass[doublecol]{epl2}
\title{Phase transitions in dependence of apex predator decaying ratio in a cyclic dominant system}
\shorttitle{Phase transitions in a cyclic dominant system}

\author{D. Bazeia\inst{1}, B.F. de Oliveira\inst{2}, and A. Szolnoki\inst{3}}
\shortauthor{Bazeia, de Oliveira, and Szolnoki}
\institute{\inst{1}Departamento de F\'\i sica, Universidade Federal da Para\'\i ba, 58051-970 Jo\~ao Pessoa, PB, Brazi\\
\inst{2}Departamento de F\'\i sica, Universidade Estadual de Maring\'a, 87020-900 Maring\'a, PR, Brazil\\
\inst{3}Institute of Technical Physics and Materials Science, Centre for Energy Research, Hungarian Academy of Sciences, P.O. Box 49, H-1525 Budapest, Hungary\\}

\abstract{Cyclic dominant systems, like rock-paper-scissors game, are frequently used to explain biodiversity in nature, where mobility, reproduction and intransitive competition are on stage to provide the coexistence of competitors. A significantly new situation emerges if we introduce an apex predator who can superior all members of the mentioned three-species system. In the latter case the evolution may terminate into three qualitatively different destinations depending on the apex predator decaying ratio $q$. In particular, the whole population goes extinct or all four species survive or only the original three-species system remains alive as we vary the control parameter. These solutions are separated by a discontinuous and a continuous phase transitions at critical $q$ values. Our results highlight that cyclic dominant competition can offer a stable way to survive even in a predator-prey-like system that can be maintained for large interval of critical parameter values.}

\begin{document}

\maketitle
Biodiversity is in general used to refer to the variety and variability of life in nature \cite{jackson_pnas75,cameron_jecol09,guill_jtb11}. It stands out as a complex phenomenon, but despite its complexity it is sometimes possible to describe features of biodiversity with the study of simplified models and the use of specific mechanisms and/or rules. Under certain circumstances, the cyclic dominance of species described by the celebrated rock-paper-scissors (RPS) model is identified as an important mechanism to maintain biodiversity
\cite{kerr_n02,kirkup_n04,lankau_s07}. As it was highlighted in a recent review \cite{szolnoki_jrsif14}, this is a rapidly developing area of evolutionary game theory where RPS-like dynamics is an intensively applied rule that can be easily extended to larger number of competing species \cite{ jiang_ll_pre11, 2012-Szolnoki-PRL-109-078701, lamouroux_pre12, avelino_pre12, avelino_pre12b, vukov_pre13, 2013-Roman-PRE-87-032148, 2014-Avelino-PRE-89-042710, szolnoki_pre14c, cheng_hy_srep14, szczesny_pre14, intoy_pre15, szolnoki_njp15, szolnoki_pre16, roman_jtb16, szolnoki_srep16b, mobilia_g16, bazeia_srep17, bazeia_epl17, brown_pre17, souza-filho_pre17, 2017-Park-SR-7-7465, 2018-Nagatani-SR-8-7094,cardinot_pa18, shibasaki_prsb18, canova_jsp18, avelino_epl18, 2018-Dobramysl-JPA-51-063001}.

Beside the above mentioned works, the mobility of competing species was also identified to promote or jeopardize diversity in intransitive competitions
\cite{reichenbach_n07}. This research path were followed by other studies \cite{jiang_ll_pre11,cheng_hy_srep14, bazeia_epl17}, where further concepts and methods were suggested. For example an interesting aspect was introduced in Refs.~\cite{bazeia_srep17,bazeia_epl17}, where the idea of Hamming distance was applied to heuristically unveil the chaotic behavior of the stochastic simulations that is usually implemented to describe biodiversity in nature. In a recent work ~\cite{intoy_pre15} Intoy and Pleimling focused on the synchronization and extinction in cyclic Lotka-Volterra models with mixed strategies and observed a transition from neutrally stable to stable when changing the level of discretization of the probability distribution. Park and his collaborators investigated the effects of nonuniform intraspecific competitions on the coexistence of species and observed that a wider spectrum of coexistence states can emerge and persist dynamically due to an alternative symmetry-breaking mechanism \cite{2017-Park-SR-7-7465}.

Notably, if we introduce a new species which dominates all previously established species then we face to a significantly new situation. The presence of this new species, that is often called as apex predator, may deeply control and/or modify the behavior of the other species in the system under consideration. In a recent work \cite{souza-filho_pre17} it was demonstrated that under specific conditions the apex individuals tend to aggregate in very small groups that spread almost uniformly in space, while the other species that compete by following the intransitive rule of the RPS game, are forming
much larger groups to keep diversity and survive in the hostile environment. 

Beyond this observation one may expect that some specific features of apex predator, including its mobility and its decaying factor, may influence relevantly the final stationary state of the global system. Motivated by this possibility, in the present work we make a systematic research to explore how the decaying ratio of apex individuals influences the vitality of other species in a RPS-type system. For this purpose we apply the model introduced in \cite{souza-filho_pre17}, but now we vary the key parameters systematically to reveal the possible destinations of evolutionary process. Furthermore, to explore and identify the emerging phase transitions we apply a previously established concept that was used to study a system where intransitive competition occurs among five species \cite{vukov_pre13}. The latter can be considered as an extension of RPS model by adding two new species, the lizard (L) and Spock (S) individuals in cyclic manner.

In the following we first describe our spatial model that is followed by the presentation of our main observations. Finally we discuss their wider implications and provide some comments for potential directions of future research.

\begin{figure}
\begin{center}
\includegraphics[width=7.0cm]{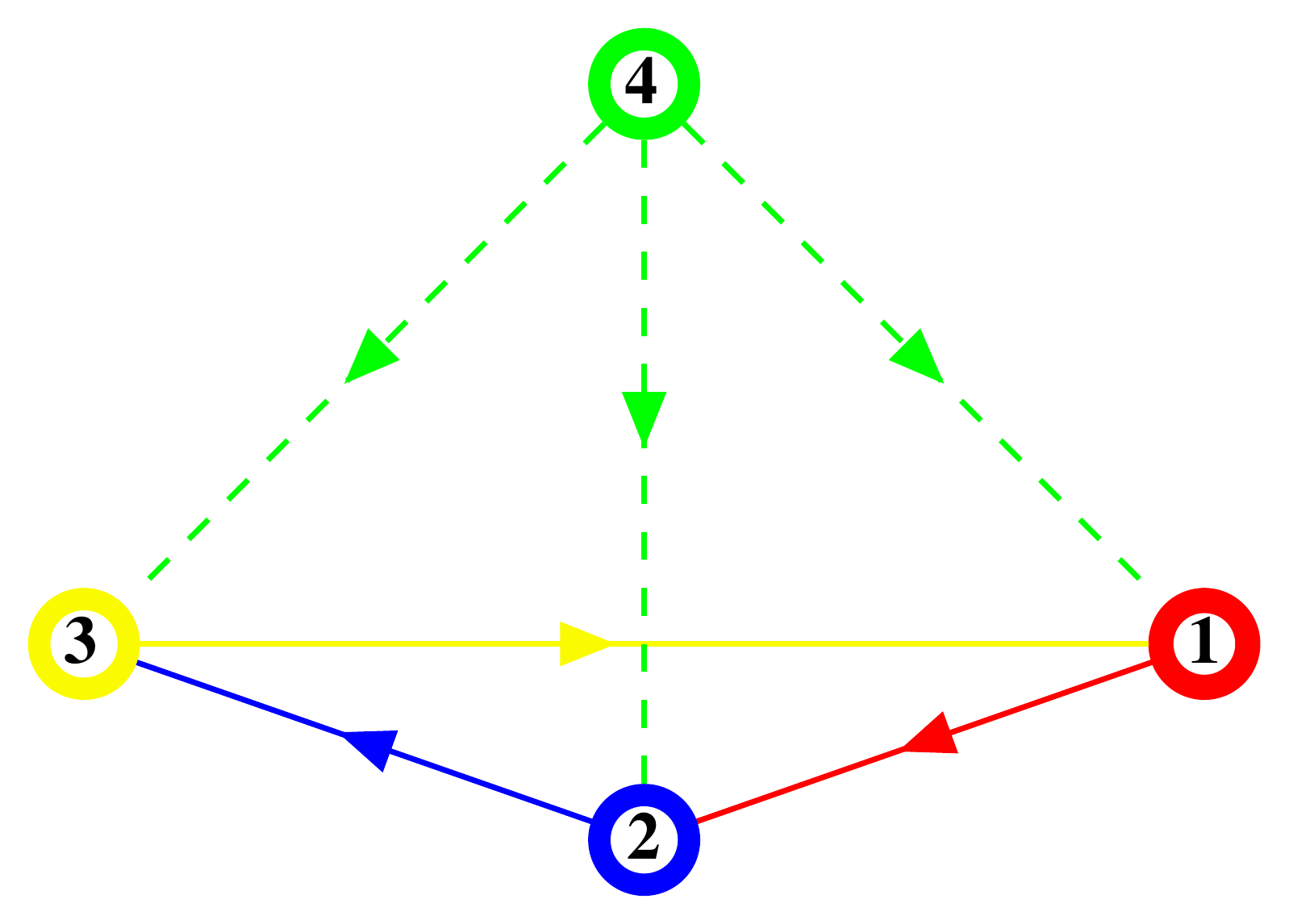}
\caption{\label{model}The diagram illustrates the schematic competition between the four species and their corresponding colors. The solid arrows occur with probability $p$, while the dashed arrows indicate the predation of the apex species with probability $a_p$.}
\end{center}
\end{figure}

Our spatial system contains three competing species which are identified by red, blue and yellow colors who are distributed on a spatial grid.
These three species interact in accordance with the RPS game, as described in Fig.~\ref{model}: red outcompetes blue, blue outcompetes yellow and yellow outcompetes red in a cyclic way, as indicated by the direction of the corresponding arrows. The invasion occurs with probability $p$, such that $p+m+r=1$, where $m$ stands for mobility and $r$ for reproduction. The cyclic invasion $p$ is described as in the May-Leonard model (see, e.g., Refs.~\cite{brown_pre17,frey_pa10}); that is, when the $p$-rule is implemented, it leaves the passive site empty, and we use the color white to indicate an empty site. Mobility describes the exchange of position between the active site and its passive neighbor while reproduction is only implemented if the passive neighbor is empty: in the latter case it is then colored by the color of the active site. 

Importantly, we introduce a fourth species, an apex predator, which is identified by green color. This new species predates all the other species but is not predated by any of them. The directions of invasions among all competitors are summarized in Fig.~\ref{model}, where the invasion strength of apex predation is described by $a_p$. The apex predation occurs as in the Lotka-Volterra model (see, e.g., Ref.~\cite{avelino_epl18}); that is, when it predates, the passive site is substituted by another apex predator. In this sense, the apex predation gives rise to the birth of apex individuals. The apex individuals can also move, and its mobility is controlled by parameter $a_m$. In order to control the apex population, they are also supposed to die, and the death is controlled by the parameter $a_d$. 

To gain comparable results with earlier studies \cite{bazeia_epl17,avelino_epl18} in this work, we consider the previously used parameter values $m=0.5$, $p=0.25$, $r=0.25$. Importantly we apply
$a_m=0.5$ and $a_p+a_d=0.5$, with $a_p=0.5/(1+q)$, where $q$ is a new parameter, that is introduced to simplify the investigation.This parameter, that can be written as $q= a_d/a_p$ and is ranged in the interval $(0,1)$, is proved to control the system behavior very efficiently. The spatial setting is described by a square lattice with size $L^2$, where the typical system size was $L=500$. By following the standard protocol we use periodic boundary conditions and the von~Neumann neighborhood, in which each site has four neighbors, located at its left, right, top and bottom side. We have done several stochastic simulations, using distinct lattice sizes and different statistical weights, but we have found similar qualitative behavior, hence the presented results can be considered as representative and remain valid in broad circumstances. 

\begin{figure*}
\begin{center}
\includegraphics[width=5.5cm]{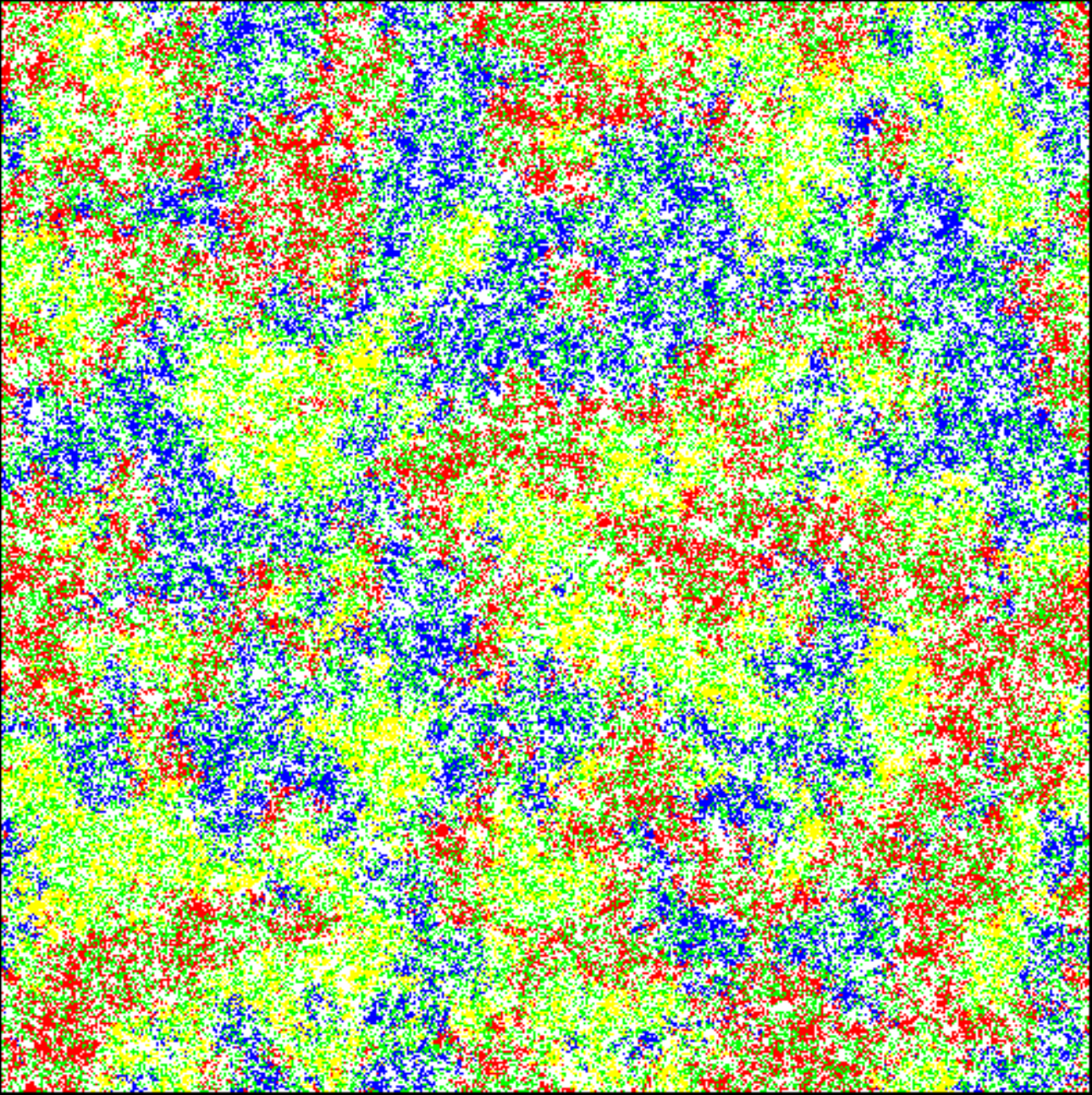}
\includegraphics[width=5.5cm]{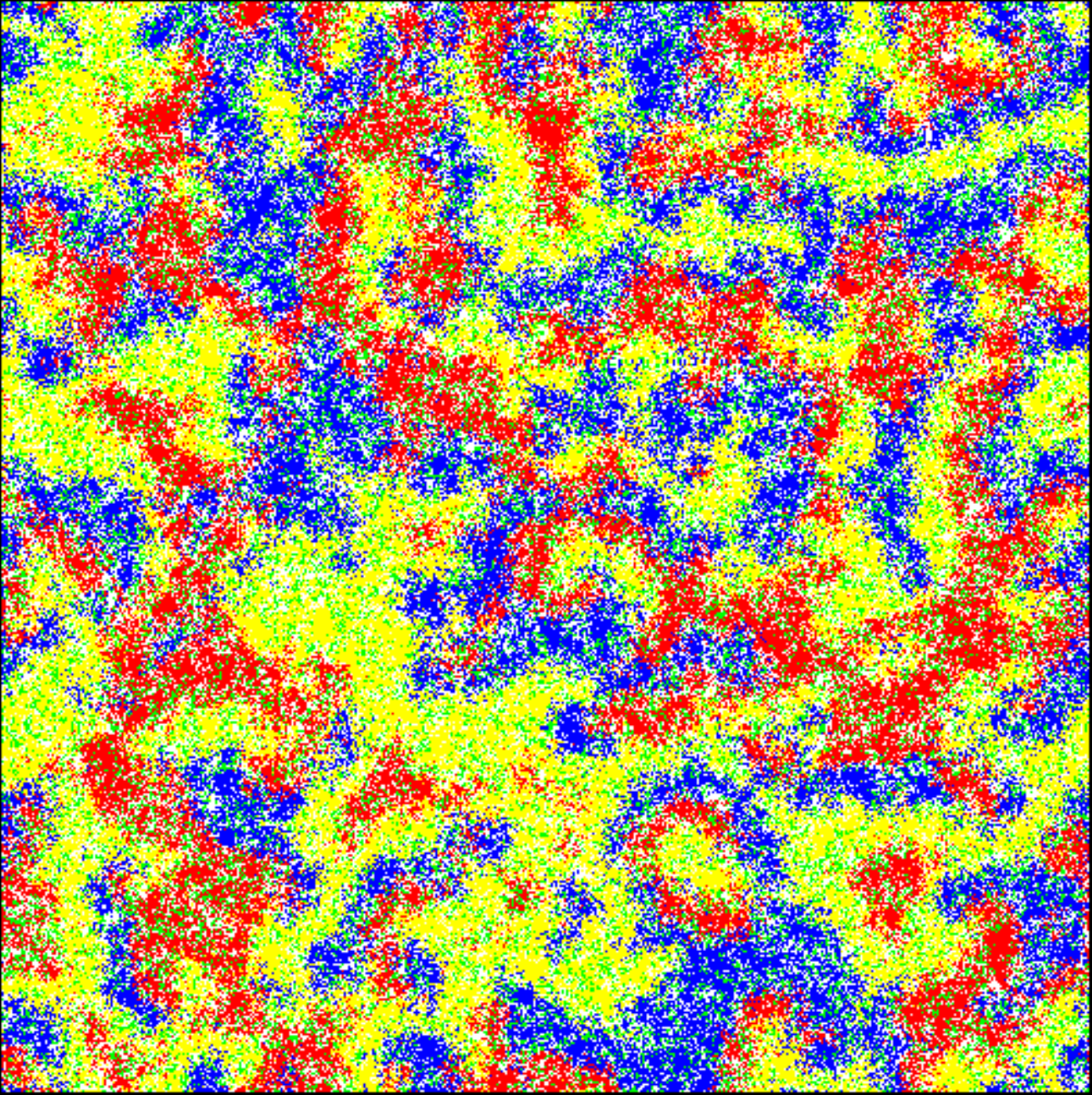}
\includegraphics[width=5.5cm]{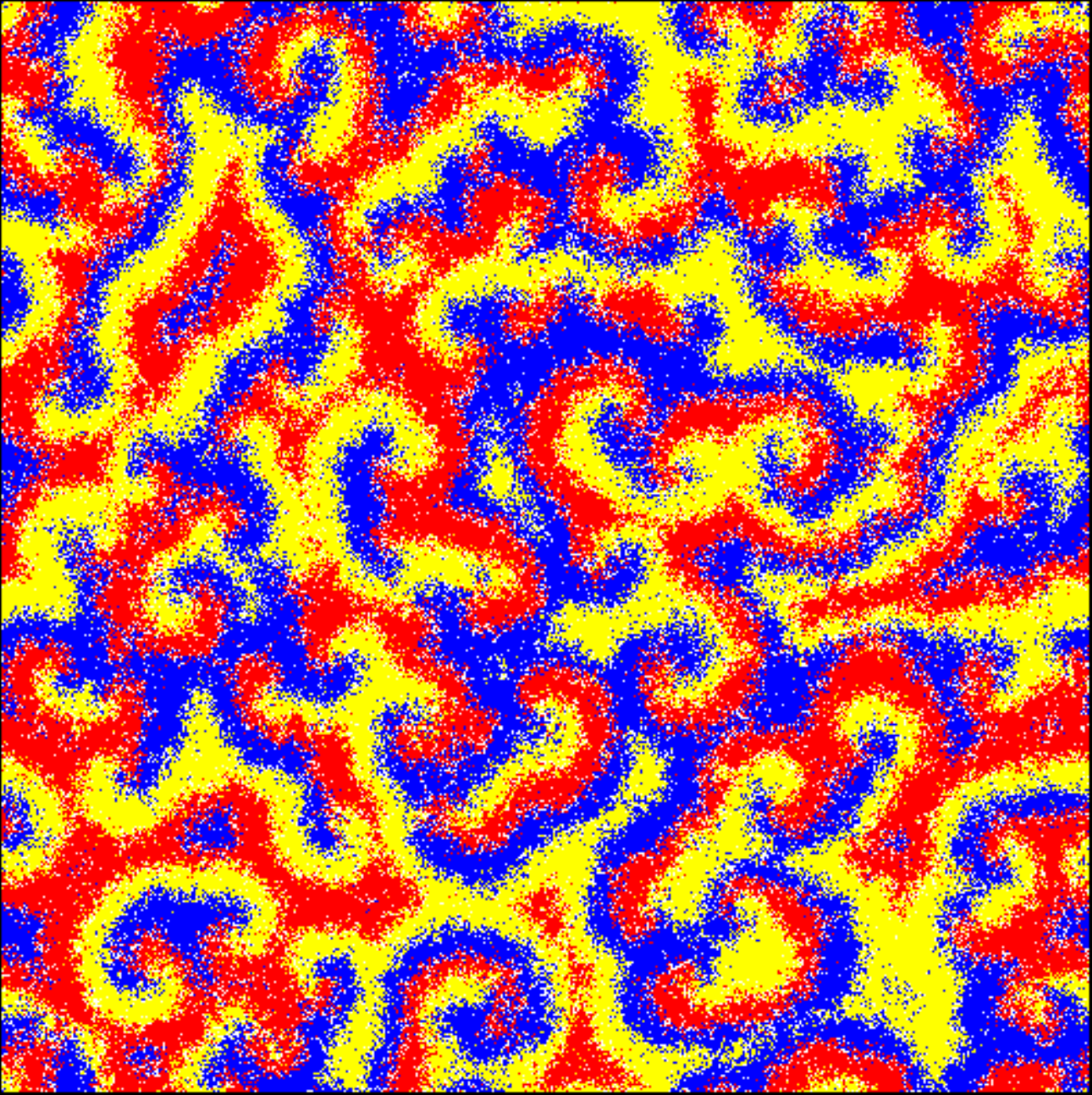}
\caption{\label{snapshots} Representative snapshots of stationary state of spatial distribution of competing species. These configurations were taken after $10000$ generations at
$q=0.28$ (left), $q=0.48$ (middle), and $q=0.68$ (right) parameter values. The linear system size is $L=500$.}
\end{center}
\end{figure*}

Before starting the time evolution, we prepare the initial state, which is constructed by first selecting a site in the lattice and then randomly painting it with the color red, blue, yellow, green or white, to represent the three competing species, the apex predator or the empty site, respectively. We repeat this $L^2$ times, to form the initial state in which the five colors are uniformly distributed in the square lattice. We use the initial state to run the stochastic simulations up to 11000 generations, with the first 1000 generations discarded. The generation time is the time spent for the system to produce $L^2$ interactions. Each interaction occurs as follows: first, one choose the active site, randomly; second, one selects the passive neighbor, also randomly. The next step is to check the active site:
\begin{itemize}
\item{if it is empty, we return and choose another active site.}
\item{if it is red, blue or yellow, the action of mobility ($m$), competition ($p$) or reproduction ($r$) is then randomly selected in accordance with the corresponding statistical weights suggested above. The action is implemented, when it is possible, and we then return to choose another active site.}
\item{if it is an apex predator, the action of mobility ($a_m$), predation ($a_p$) or death ($a_d$) is then randomly selected in accordance with the corresponding statistical weights suggested above. The action is implemented, when it is possible, and we then return to choose another active site.}
\end{itemize}

The parameter $q$, which will be used below, is inspired by Ref.~\cite{vukov_pre13}, where an investigation of the fluctuations of the five species densities (the RPSLS model) revealed a diverging behavior at the critical point. However, although in the current model we are also dealing with five distinct possibilities, with the colors red, blue, yellow, green and white, one notes that they do not evolve at the same footing, since the color white represents an empty (passive) site. Furthermore, the color green describes the apex predator that has its own statistical rules, different from the rules of the subset with the colors red, blue and yellow that evolve in accordance with the RPS model. In this sense, we cannot follow the lines of Ref.~\cite{vukov_pre13} directly, but we focus mainly on the abundance of the apex predator and investigate its role on the system behavior in dependence of the control parameter $q$. 

To provide a general overview about the system behavior on different values of control parameter in Fig.~\ref{snapshots} we display representative snapshots of the stationary state after 10000 generations, for three distinct values of $q$. These plots suggest that the density of apex predators decays as $q$ increases and above a critical $q$ value there are no apex individuals anymore. The latter is shown in the last panel, obtained at $q=0.68$, where three species red, blue and yellow develop the standard pattern of spirals characterizes three distinct species that evolve under the rules of the RPS competition game in a spatial population. We have also checked that the patterns displayed in Fig.~\ref{snapshots} keep the same qualitatively features for longer and longer times, so they are typical representations of the system for the three specific values of $q$. In this sense, these snapshots illustrate the presence of diversity under the action of the apex predator in the first two panels, and after its extinction, in the right panel.

\begin{figure}
\begin{center}
\includegraphics[width=8.0cm]{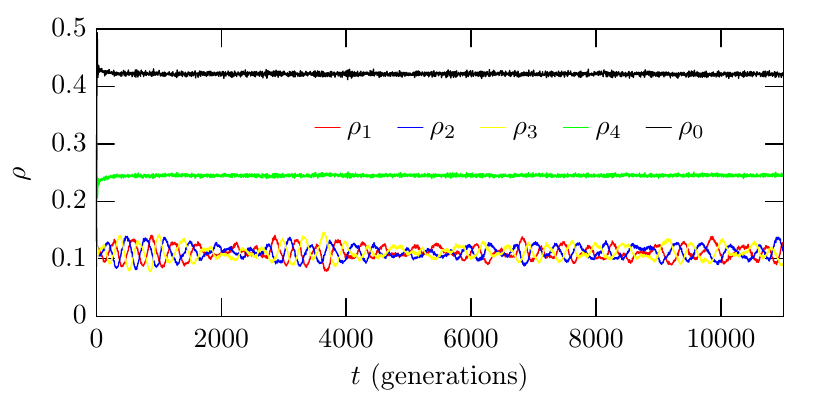}\\
\includegraphics[width=8.0cm]{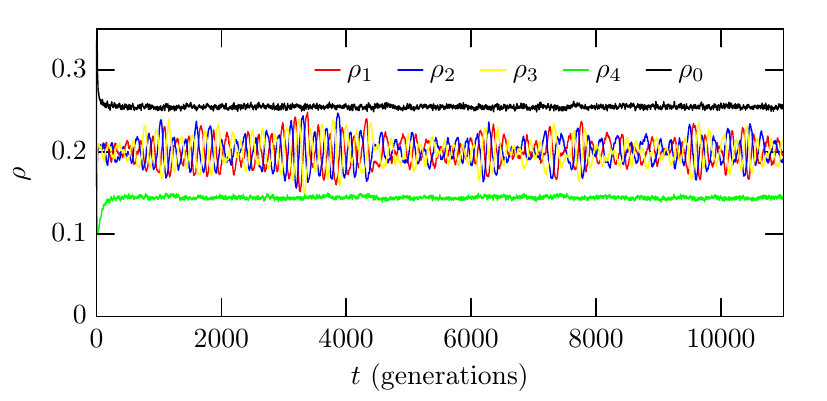}\\
\includegraphics[width=8.0cm]{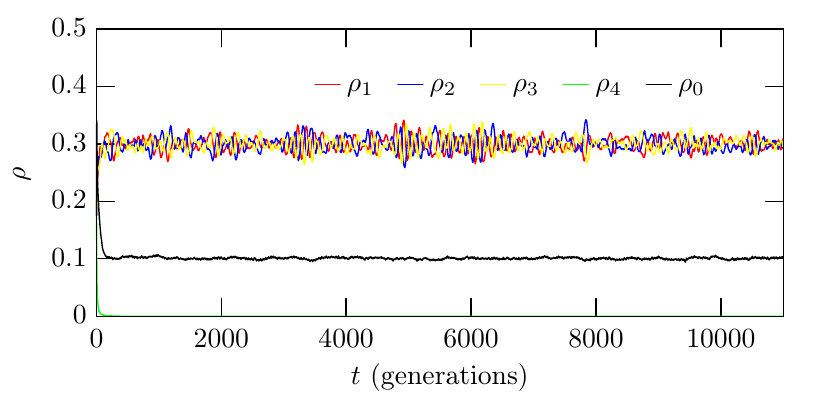}
\caption{\label{time} The abundance or density $\rho_j(t)$ is depicted as a function of time, for three representative values of $q$. We display the cases with $q=0.28$ (top), $q=0.48$ (middle), and $q=0.68$ (bottom).}
\end{center}
\end{figure}

To gain a deeper quantitative insight about the time evolution in Fig.~\ref{time} we depict the abundance or density, that is, the number of individuals divided by $L^2$ for all species. The results show that the abundance of red, blue and yellow individuals oscillate around the same average, which depends on the abundance of the apex predator. They also show that the abundance of the apex predator diminishes as $q$ increases, and that it is zero for $q$ at the value $0.68$. The abundance of the empty sites is shown in black, and it also decreases as $q$ increases; however, it does not vanish, reaching the lowest value around $0.1$, in the absence of the apex predator.

\begin{figure}
\begin{center}
\includegraphics[width=8.0cm]{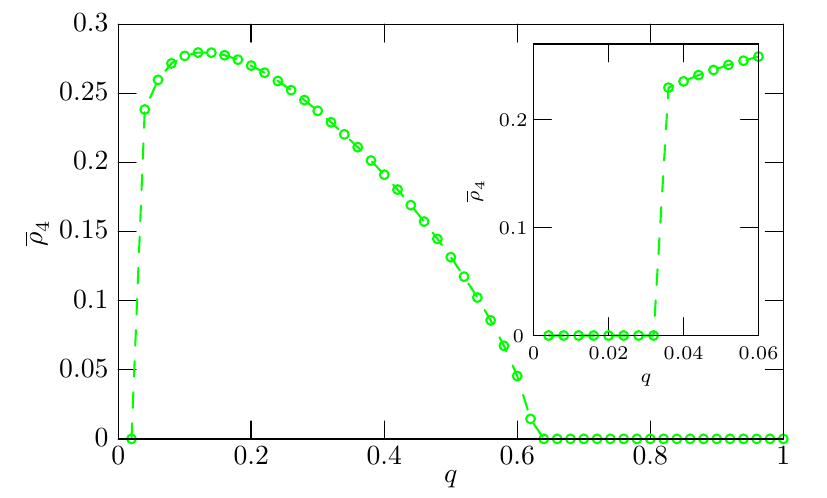}
\caption{\label{comprehensive} Comprehensive view about the average density of the apex individuals, $\bar\rho_4=\langle\rho_4(t)\rangle$, which is depicted as a function of $q$, for a single realization. The inset shows the discontinuous phase transition at $q_{c1}=0.035(1)$. The continuous phase transition is at $q_{c2}=0.62(1)$. The error bars are smaller than the size of symbols.}
\end{center}
\end{figure}

The results displayed in Figs.~\ref{snapshots} and \ref{time} are compatible with one another, and they suggest that the system undergoes a phase transition as $q$ increases, somewhere in between $0.48$ and $0.68$. For this reason, we decide to investigate the average number of individuals as we vary $q$. We call this $\bar\rho_j=\langle\rho_j(t)\rangle$, with $j=1,2,3,4,$ and $0$, standing for red, blue, yellow, green, and white, representing the three species, the apex predator and the empty sites, respectively. 
Note that we are principally interested in the influence of the apex predator on the three-species model, therefore we display the average density of the apex predator in Fig.~\ref{comprehensive} for the whole interval of control parameter. The result suggests that there are two transitions as we vary $q$. The first is in the $q \le 0.1$ interval, where a finite discontinuity appears suggesting a first-order phase transition, and the second is in the other interval, with $q \ge 0.1$ where $\bar \rho_4$ also goes to zero, but now continuously.

In the case of small $q$, we concentrated on the first order phase transition and investigated the interval $q \in (0,\,0.06)$ carefully. The results are displayed in the inset of Fig.~\ref{comprehensive}, and it clearly shows the presence of a first order phase transition at the critical value $q_{c1}=0.035(1)$. To obtain $q_{c1}$ with the desired accuracy, it was calculated from an average over 100 realizations, each one starting from a different initial state. This phase transition indicates that when $q$ is less than $0.1$ and diminishes towards $q_{c1}$, the apex predator starts becoming too dominant, diminishing the red, blue and yellow species in a way such that they cannot provide enough food for the predator anymore. As a result the full system collapses and all species go extinct.

\begin{figure}
\begin{center}
\includegraphics[width=8.0cm]{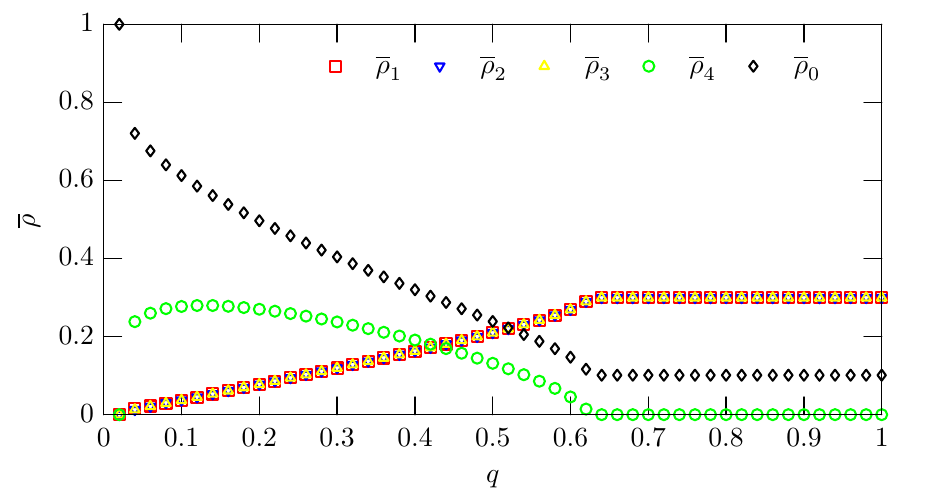}
\caption{\label{continuous} The average density $\bar\rho_j=\langle\rho_j(t)\rangle$ is depicted as a function of $q$. The value $q_{c2}=0.62(1)$ shows where apex predators go extinct and the system terminates onto the traditional rock-scissors-paper game-type solution. The error bars are smaller than the size of symbols.}
\end{center}
\end{figure}

The other phase transition is detected for larger values of $q$. Here we keep using $L=500$ as the system size, which is proved to be large enough to reach the desired accuracy. We first collect results for the average density $\bar\rho_j=\langle\rho_j(t)\rangle$, with $j=1,2,3,4,$ and $0$, standing for red, blue, yellow, green, and white, representing the three species, the apex predator and the empty sites, respectively. These results are displayed in Fig~\ref{continuous} and there we see that $\bar \rho_4$ vanishes continuously as $q$ increases, confirming the results of Fig.~\ref{comprehensive} and suggesting that the system undergoes another phase transition at the second critical value $q_{c2}=0.62(1)$. The results depicted in Fig.~\ref{continuous} are also compatible with the ones displayed in Fig.~\ref{snapshots} and~\ref{time}.

\begin{figure}
\begin{center}
\includegraphics[width=8.0cm]{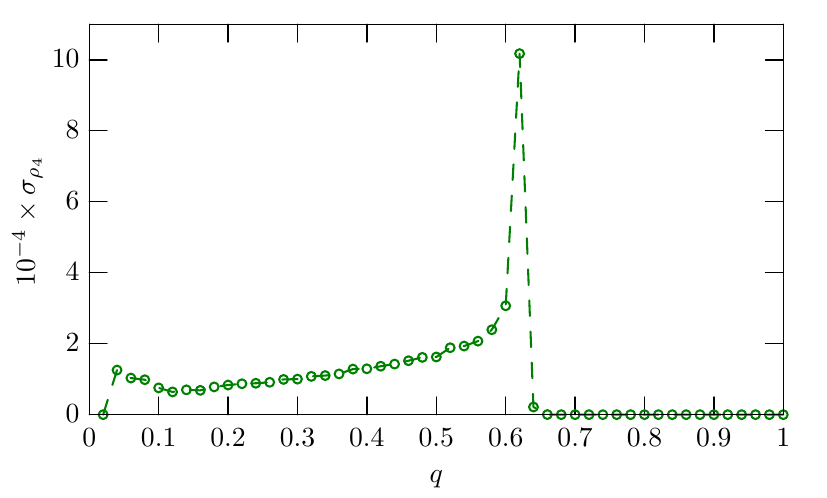}
\caption{\label{deviation} The standard deviation $\sigma_{\rho_4}(q)$ of the apex predator is depicted as a function of $q$. It shows the appearance of a discontinuity that occurs at the value $q_{c2}=0.62(1)$, indicating the presence of a second-order phase transition of the order parameter. The error bars are smaller than the size of symbols.}
\end{center}
\end{figure}

In order to better understand the continuous phase transition that we have just identified, at $q_{c2}=0.62(1)$, we apply the previously recommended method \cite{vukov_pre13} and study the standard deviation of the order parameter of the apex predator, which we call $\sigma_{\rho_4}(q)$. The results are displayed in Fig.~\ref{deviation}, and there we can see clearly the emergence of a discontinuity at the value $q_c=0.62(1)$, which also confirms our previous results, displayed in Figs.~\ref{comprehensive} and \ref{continuous}. But now it shows that the system undergoes a second order phase transition, since the discontinuity is related to the fluctuations of the order parameter that is taken to be the standard deviation $\sigma_{\rho_4}(q)$; for further details see, e.g., Ref.~\cite{reichl_09}. In Fig.~\ref{deviation}, we have depicted the results after an average over 50 realizations, each one starting with an independent initial state. The issue here is that as the parameter $q$ increases towards $q_{c2}$, the apex individuals diminish smoothly until they are fully extinguished, keeping  the red, blue and yellow species evolving with the standard spiral patterns typical of the cyclic dominant three-species system.         

The stochastic simulations that we have implemented in this work are always based on random choices, by selecting an active site, a passive neighbor and the rule, and then implementing the action. In this sense, the action selected is not always implemented, because it depends on the active site, the passive neighbor and the specific rule. For instance, if the rule is selected for
reproduction but the chosen passive neighbor is not empty, no action is implemented. This has led us to think about the possibility to study how the number of actions effectively implemented during the time evolution of the system depends on $q$. Considering it as a density, we called it occurrence rate, which is represented by the letter $n$. It adds positively, unveiling an independent way to detect a phase transition and more importantly, exposing another interesting feature related to the three-species model.

\begin{figure}[b!]
\begin{center}
\includegraphics[width=8.0cm]{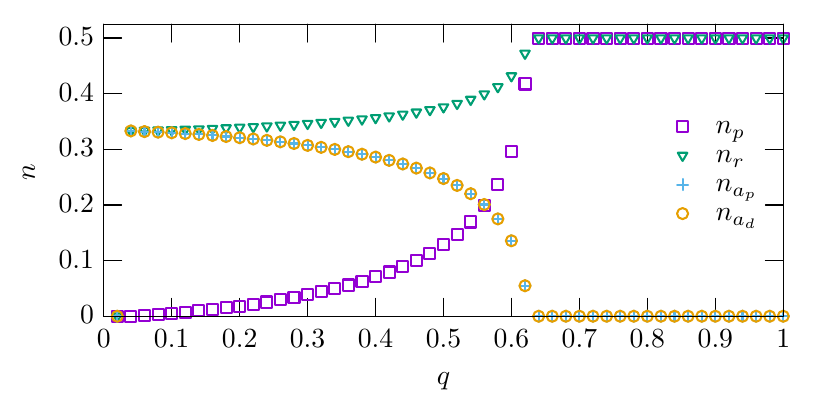}
\caption{\label{actions} The density of actions effectively implemented by the stochastic simulations is displayed as function of $q$. The results clearly suggest a continuous phase transition in dependence of $q$ at $q_{c2}=0.62(1)$. The error bars are smaller than the size of symbols.}
\end{center}
\end{figure}

The results are displayed in Fig.~\ref{actions}. They are obtained from the average over 50 realizations, each one starting with an independent initial state, with the lattice size $L=500$. There we display $n_p$ and $n_r$, which represent the occurrence rate related to the rules competition and reproduction for the red, blue and yellow individuals, and also $n_{a_p}$ and $n_{a_d}$, which are similar quantities related to the apex predator. The results confirm that the phase transition occurs at $q_c=0.62(1)$, but they also show other interesting features of the system. One notes, for instance, that in the presence of the apex predator, $n_p$ is always below $n_r$, until the extermination of the predator, where they equate each other to keep biodiversity. This is an interesting result, and it informs us that in the presence of the apex predator, reproduction occurs at a rate always above competition. Moreover, both $n_p$ and $n_r$ increase as $q$ increases, following different ratios to reach the same value at the phase transition. On the other hand, the time evolution keeps $a_p=a_d$, showing that the average number of deaths equals the average number of births, and also that both numbers decrease as $q$ increases. These results constitute other distinct tests to confirm the second order phase transition that we have already identified. We further note that the equality between $n_{a_p}$ and $n_{a_d}$ that appear in Fig.~\ref{actions} nicely conforms with the result displayed in Fig.~\ref{time}, which shows that the population of apex individuals is almost constant in time. 

Summing up, in this work we have investigated a system composed of three distinct species (red, blue, and yellow) that evolve competing cyclically under the rules of the rock-paper-scissors game. These species interact under the presence of an apex predator, identified by the color green, that predate all the three species red, blue and yellow, but is not predated by any of them. We have studied how the three species evolve controlled by the the rules of the apex predator, to see the importance of the apex to control the subset of species of the RPS model. 

We have implemented several distinct stochastic simulations and identified two qualitatively different phase transition points in dependence of the control parameter $q$. The first one is a discontinuous phase transition below which the  whole system goes extinct due to the greedines of apex predator. The two phase transitions can be described by the order parameter $\bar \rho_4$, that represents the abundance of the apex predator. This order parameter depends on $q$, that is, $\bar \rho_4= \bar\rho_4(q)$, with $q$ standing for the ratio between the two parameter $a_d$ and $a_p$ that control the apex death and predation weights, respectively. The first order phase transition occurs at $q_{c1}=0.035(1)$, which is the critical value where the order parameter $\rho_4(q)$ engenders a discontinuous behavior. The other phase transition occurs at $q_{c2}=0.62(1)$, which is the critical value where $\rho_4(q)$ vanishes continuously. In the latter case, we have also studied the standard deviation
$\sigma_{\rho_4}(q)$, related to order parameter $\bar \rho_4(q)$ of the apex individuals. The results show that when the death weight of the apex predator increases and parameter $q$ reaches the critical value $q_c=0.62(1)$, the apex individuals are extinguished and the three species red, blue and yellow keep evolving in the standard way, exhibiting the spiral patterns that are typical for spatial RPS model.

In this sense, we have found a good variable and an effective order parameter, from which we could distinguish two distinct phase transitions in the system. One is the continuous phase transition in which the apex predator disappears, leaving the other three species evolving with the spiral pattern typical of the RPS model. In the presence of the apex predator, we also noted that in the three-species model, competition occurs at a rate always below reproduction. Our observation emphasizes the particular robustness of cyclic dominant coexistence. More precisely, such solution can survive even at harsh environment until the superior apex predator becomes too dominant. In the latter case the average density of cyclic dominant species becomes so small that cannot provide enough food for apex predator and the whole system collapses.

The above results are of current interest and can be used to investigate several other problems, in particular the case with a greater number of species and other rules to control their collective behavior. Furthermore, our observations revealed that in the presence of a predator the three cyclically competing prey species experience higher growth rates, and lower competitive death rates, than they do comparing to the classic model in the absence of apex predator. This offers a novel mechanism of predator-mediated biodiversity enhancement and we do hope that it will stimulate further interesting works on this field \cite{park_c18, szolnoki_njp18, ichinose_njp18, park_c18c}.

There are many possibilities, and we are now examining some specific cases, hoping to report on them in the near future. In particular, the observation that the three competing species experience higher growth rates and lower competitive death rates in the presence of the predator suggests a potential mechanism of predator-mediated biodiversity enhancement. Another specific question concerns the study of the critical values of the parameter $q$ as a function of the mobility of the apex predator. Furthermore, since the cooperator-punisher-defector system \cite{r1, r2} is in the cyclic dominant class of systems, it would also be of interest to examine it under similar conditions. These are interesting issues which will be further investigated elsewhere.

\begin{acknowledgments}
This research was supported by the Brazilian agencies CAPES, CNPq, Funda\c c\~ao Arauc\'aria, INCT-FCx and by the Hungarian National Research Fund (Grant K-120785)
\end{acknowledgments}

\end{document}